\documentclass[12pt]{article} 
  \usepackage{epsfig} 
  \usepackage{psfrag} 
  \usepackage{a4} 
 
  \newcommand{\be}{\begin{equation}} 
  \newcommand{\ee}{\end{equation}} 
 
\begin{document}

  \title{Introducing Variety in Risk Management} 
 
  \author{Fabrizio Lillo, Rosario N. Mantegna,\\  
  Jean-Philippe Bouchaud  and Marc Potters} 
  \date{}
  \maketitle

  ``Yesterday the S\&P500 went up by 3\%''. Is this number telling  
  all the story if half the stocks went up 5\% and half went down  
  1\%? Surely one can do a little better and give  
  {\it two} figures, the average and the dispersion around  
  this average, that two of us have recently christened  
  the {\it variety} \cite{Lillo-Mantegna2000}.  
 
  Call $r_i(t)$ the return of asset $i$ on day $t$. The variety  
  ${\cal V}(t)$ is simply the root mean square of the stock 
  returns on a given day: 
  \be 
  {\cal V}^2(t) = \frac{1}{N}\sum_{i=1}^{N} 
  \left(r_i(t)-r_m (t)\right)^2,\label{variety} 
  \ee 
  where $N$ is the number of stocks and $r_m = (1/N) \sum_i r_i$ 
  is the market average. 
  If the variety is, say, 0.1\%, then most stocks have indeed made between 
  2.9\% and 3.1\%. But if the variety is 10\%, then stocks followed 
  rather different trends during the day and their average 
  happened to be positive, but this is just an average information. 
 
  The variety is {\it not} the volatility of the index. The volatility 
  refers to the amplitude of the fluctuations of the index from one day 
  to the next, not the dispersion of the result {\it between different 
  stocks}. Consider a day where the market has gone down 5\% with a 
  variety of 0.1\% -- that is, all stocks have gone down by nearly 5\%. 
  This is 
  a very volatile day, but with a low variety. Note that low variety 
  means that it is hard to diversify: all stocks behave the same way. 
 
  The intuition is however that there should be a correlation between 
  volatility and variety, probably a positive one: when the market makes 
  big swings, stocks are expected to be all over the place. This is 
  actually true. Indeed the correlation coefficient between ${\cal V}(t)$ 
  and $|r_m|$ is 0.68.\footnote{This value is not an artifact due to  
  outliers. In fact an estimation of the Spearman rank-order  
  correlation coefficient gives the value of 0.37 with a significance level 
  of $10^{-35}$.} 
  The variety is, on average, larger when the amplitude  
  of the market return is larger (see the discussion below). Very much  
  like the volatility, the variety is correlated in time:  
  there are long periods 
  where the market volatility is high and where the market variety is 
  high (see Fig.\ 1).  Technically, the temporal correlation 
  function of these two objects reveal a similar slow (power-law like)  
  decay with time.

  \vspace{1cm} 
  \centerline{\fbox{Figure 1 about here}} 
  \vspace{1cm}

  A theoretical relation between variety and market average return  
  can be obtained within the framework of the one-factor 
  model, that suggests that the variety increases when the market 
  volatility increases. 
  The one-factor model assumes that $r_i(t)$ 
  can be written as: 
  \be 
  r_i(t) = \alpha_i + \beta_i {\cal R}_m(t) + \epsilon_i(t),\label{eq2} 
  \ee    
  where $\alpha_i$ is the expected value of the component  
  of security $i$'s return that is independent of the market's 
  performance (this parameter usually plays a minor role and 
  we shall neglect it), $\beta_i$ is a coefficient usually close to unity 
  that we will assume to be {\it time independent}, 
  ${\cal R}_m(t)$ is the market factor and 
  $\epsilon_i(t)$ is called the idiosyncratic return, by construction 
  uncorrelated both with the market and with other idiosyncratic factors. 
  Note that in the standard one-factor model the distributions 
  of ${\cal R}_m$ and $\epsilon_i$ are chosen to be Gaussian with constant 
  variances, we do not make this assumption and let these 
  distributions be completely general including possible volatility  
  fluctuations. 
  
  In the study of the properties of the one-factor model 
  it is useful to consider the variety $v(t)$ of idiosyncratic part, 
  defined as 
  \be 
  v^2(t) ={\frac{1}{N} \sum_{i=1}^N [\epsilon_i(t)]^2}. 
  \ee 
  Under the above assumptions the relation between the variety  
  and the market average return is well approximated by (see Box 1 
  for details): 
  \be 
  {\cal V}^2(t) \simeq v^2(t) + \Delta \beta^2 r_m^2(t),\label{eq4} 
  \ee 
  where $\Delta \beta^2$ is the variance of the $\beta$'s divided 
  by the square of their mean.  
 
  Therefore, even if the idiosyncratic variety $v$ is constant, Eq. (4) 
  predicts an increase of the volatility with $r_m^2$, which is a 
  proxy of the market volatility. Because $\Delta \beta^2$ is small, however,  
  this increase is rather small. As we shall now discuss, the effect is 
  enhanced by the fact that $v$ itself increases with the market volatility.  
 
  In its simplest version, the one-factor model assumes 
  that the idiosyncratic part $\epsilon_i$ is {\it independent} of the 
  market return. In this case, the variety of idiosyncratic  
  terms $v(t)$ is constant in time and independent from $r_m$. In Fig. 2 we 
  show the variety of idiosyncratic terms as a function of the market  
  return. In contrast with these predictions, the empirical  
  results show that a significant correlation  
  between $v(t)$ and $r_m(t)$ indeed exists. The degree of correlation is 
  different for positive and negative values of the market average.   
  In fact, the best linear least-squares fit between $v(t)$ and  
  $r_m(t)$ provides different slopes when the fit is performed for positive  
  (slope $+0.55$) or negative (slope $-0.30$) value of the market average. 
  We have again checked that these slopes are not governed by outliers 
  by repeating the fitting procedure in a robust way. 
  The best fits obtained with this procedure are shown in Fig. 2 as 
  dashed lines. The slopes of the two lines are -0.25 and 0.51 
  for negative and positive value of the market average, respectively.    
  Therefore, from Eq.(4) we find that the increase of variety in  
  highly volatile periods is stronger than what is expected from the  
  simplest one-factor model, although not as strong for negative (crashes) 
  than it is for positive (rally) days. 
  By analyzing the three largest crashes occurred at the NYSE in the 
  period from January 1987 to December 1998, we observe two 
  characteristics of the variety which are recurrent during the  
  investigated crashes: (i) the variety increases substantially  
  starting from the crash day 
  and remains at a level higher than typical for a period of time 
  of the order of sixty trading days; (ii) the highest value of the  
  variety is observed the trading day immediately after the crash.

  \vspace{1cm} 
  \centerline{\fbox{Figure 2 about here}} 
  \vspace{1cm} 
 
  An important quantity for risk management purposes is the 
  degree of correlation between stocks. If this correlation is 
  too high, diversification of risk becomes very difficult to achieve. 
  A natural way \cite{QF} to characterize the average  
  correlation between all stocks $i,j$ on a given day  ${\cal C}(t)$ 
  is to define the following quantity: 
  \be 
  {\cal C}(t) = \frac{\frac{1}{N(N-1)}\sum_{i \neq j} r_i(t) 
  r_j(t)}{\frac{1}{N} \sum_{i} r_i^2(t)}.\label{eq5} 
  \ee  
  As shown in the technical Box 1, to a good approximation one finds: 
  \be 
  {\cal C}(t) \simeq 
  \frac{1}{1+F(t)},  
  \qquad F(t) \equiv \frac{v^2(t)}{r_m^2(t)},\label{eq6} 
  \ee  
 
  As mentioned above, the variety of the idiosyncratic terms is {\it 
  constant}  
  in time in the simplest one-factor model.  The 
  correlation structure in this version of the one-factor model is very 
  simple and time independent. Still, the quantity ${\cal C}$, taken for 
  a proxy of the correlations on a given day, increases with the 
  `volatility' $r_m^2$, simply because $F$ decreases. As shown in Fig.\ 3, the 
  simplest one factor model in fact overestimates this increase \cite{QF}.   
  Because the idiosyncratic variety $v^2(t)$ tends to increase when $|r_m|$  
  increases (see Figure 2),  
  the quantity $F(t)$ is in fact larger and ${\cal C}$ is smaller.  
  This may suggest that, at odds with the common lore,  
  correlations actually are less effective than expected  
  using a one-factor model in high volatility periods: the unexpected 
  increase of variety gives an additional opportunity for diversification. 
  Other, more subtle 
  indicators of correlations, like the exceedance correlation function 
  defined in Box 2 and shown in Fig.\ 4 (see \cite{Longin}), actually confirm that  
  the commonly reported 
  increase of correlations during highly volatile bear periods might 
  only reflect the inadequacy of the indicators that are used to measure them. 
 
  \vspace{1cm} 
  \centerline{\fbox{Figures 3 and 4 about here}} 
  \vspace{1cm} 
 
  Therefore, the idiosyncrasies are by construction uncorrelated, but not 
  independent of 
  the market. This shows up in the variety, does it also appear in 
  different quantities? We have proposed above to add to the market 
  return the variety as a second indicator. One can probably handle 
  a third one, which gives a refined information of what happened in the 
  market on a particular day.  The natural question is indeed: what 
  fraction $f$ of stocks did actually better than the market?  A 
  balanced market would have $f=50\%$. If $f$ is larger than $50\%$, 
  then the majority of the stocks beat the market, but a few ones lagging 
  behind rather badly, and vice versa. A closely related measure is the 
  {\it asymmetry} ${\cal A}$, defined as ${\cal A}(t) = r_m(t) - r^*(t)$,  
  where the median $r^*$ is, by definition, the return such that 
  50\% of the stocks are above, 50\% below. If $f$ is larger than 50\%, 
  then the median is larger than the average, and vice versa. Is the 
  asymmetry ${\cal A}$ also correlated with the market factor? Fig.\ 5 
  shows that it is 
  indeed the case: large positive days show a positive skewness in the 
  distribution 
  of returns -- that is, a few stocks do exceptionally well -- whereas 
  large negative days show the opposite behaviour. In the figure each day 
  is represented by a circle and all the circles cluster in a  
  pattern which has a sigmoidal shape. The asymmetrical behaviour observed 
  during two extreme market events is shown in the insets of Fig. 5 where  
  we present the probability density function of returns observed in the 
  most  
  extreme trading days of the period investigated in  
  Ref. \cite{Lillo-Mantegna2000b}.  
  This empirical observation cannot be explained by a one-factor  
  model. This has been shown by two different approaches: (i) 
  by comparing empirical results with surrogate data generated by 
  a one-factor model \cite{Lillo-Mantegna2000b} and (ii) by 
  considering directly the asymmetry of daily idiosyncrasies \cite{QF}.     
  Intuitively, one possible explanation of this anomalous skewness  
  (and a corresponding increase of variety) might be related to the existence  
  of sectors which strongly separate from each other during volatile days. 
 
  \vspace{1cm} 
  \centerline{\fbox{Figure 5 about here}} 
  \vspace{1cm}

  The above remarks on the dynamics of stocks seen as a population are 
  important for risk control, in particular for option books, and for 
  long-short equity trading programs.  The variety is in these cases 
  almost as important to monitor as the volatility. Since this quantity 
  has a very intuitive interpretation and an unambiguous definition 
  (given by Eq.\ (1)), this could 
  become a liquid financial instrument which may be used to hedge 
  market neutral positions. Indeed, market neutrality is usually  
  insured for `typical' days, but is destroyed in high variety 
  days. Buying the variety would in this case reduce the risk of these 
  approximate market neutral portfolios. 
  \vskip 1cm 
   
  {\bf Fabrizio Lillo and Rosario N. Mantegna are with the Observatory 
  of Complex Systems, a research group of Istituto Nazionale per la 
  Fisica della Materia, Unit of Palermo and Dipartimento di Fisica e 
  Tecnologie Relative of Palermo University, Palermo 
  Italy. Jean-Philippe Bouchaud and Marc Potters are at Science \& 
  Finance, the research division of Capital Fund 
  Management. Jean-Philippe Bouchaud is also at the Service de 
  Physique de l'Etat Condens\'e, CEA Saclay.} 
 
  \section{Technical Box 1: Proof of Eqs. (4) and (6)} 
 
  Here we show that if the number of stocks $N$ is large, then up to terms 
  of order $1/\sqrt{N}$, Eqs. (\ref{eq4}) and (\ref{eq6}) indeed hold. We  
  start from Eq. (\ref{eq2}) with $\alpha_i \equiv 0$. Summing over $i=1,..,N$ 
  this equation, we find: 
  \be 
  r_m(t) = {\cal R}_m(t) \frac{1}{N}\sum_{i=1}^N \beta_i  + \frac{1}{N}\sum_{i=  1}^N \ 
  \epsilon_i(t) 
  \ee 
  Since for a given $t$ the idiosyncratic factors are uncorrelated from  
  stock to stock, the second term on the right hand side is of order  
  $1/\sqrt{N}$, and can thus be neglected in a first approximation 
  giving  
  \be 
  r_m(t) \simeq \overline \beta {\cal R}_m(t), 
  \ee 
  where  
  $\overline \beta \equiv \sum_{i=1}^N \beta_i/N$. In order to obtain Eq. (4), 
  we square Eq. (2) and summing over $i=1,..,N$, we find:  
  \be 
  \frac{1}{N}\sum_{i=1}^N r_i^2(t) =  
  {\cal R}_m^2(t) \frac{1}{N}\sum_{i=1}^N \beta_i^2  +  
  \frac{1}{N}\sum_{i=1}^N \epsilon_i^2(t)+ 
  2 ~ {\cal R}_m(t) \frac{1}{N}\sum_{i=1}^N \beta_i \epsilon_i(t) 
  \ee 
  Under the assumption that $\epsilon_i(t)$ and $\beta_i$ are  
  uncorrelated the last term can be neglected and the variety 
  ${\cal V}(t)$ defined in Eq. (1) is given by 
  \be 
  {\cal V}^2(t) \simeq v^2(t) + (\overline {\beta^2} -\overline \beta^2) {\cal R}_m^2(t) 
  \ee   
  This is the relation between ${\cal V}(t)$ and the market factor  
  ${\cal R}_m(t)$. By inserting Eq. (8) in the previous equation 
  one obtains Eq. (4).   
   
  Now consider Eq. (\ref{eq5}). Using the fact that $\sum_{i=1}^N r_i/N =r_m$, 
  we find that the numerator is equal to $r_m^2$ up to terms of order $1/N$.  
  Inserting Eq. (\ref{eq2}) in the denominator and again neglecting the cross-product 
  terms that are of order $1/\sqrt{N}$, we find: 
  \be 
  {\cal C}(t) \simeq \frac{r_m^2(t)}{\overline{\beta^2} {\cal R}_m^2(t) 
  +v^2(t)} \simeq \frac{1}{\chi+F(t)},  
  \qquad F(t) \equiv \frac{v^2(t)}{r_m^2(t)},\label{eq6bis} 
  \ee 
  where $\chi \equiv \overline {\beta^2} /\overline \beta^2$.  
  This quantity is empirically  
  found to be $\simeq 1.05$ for the S\&P 500,  
  and we have therefore replaced it by $1$  
  in Eq. (\ref{eq6}). 
 
  \section{Box 2: Exceedance correlations} 
 
  In order to test the structure of the cross-correlations during highly  
  volatile periods, Longin and Solnik (\cite{Longin}) have proposed to study  
  the `exceedance correlation', defined for for a given pair $ij$ of stocks as follows: 
\be 
\rho_{ij}^+(\theta)=\frac{\langle \tilde r_i \tilde r_j \rangle_{>\theta} 
- \langle \tilde r_i \rangle_{>\theta} \langle \tilde r_j \rangle_{>\theta}} 
{\sqrt{(\langle \tilde r_i^2 \rangle_{>\theta} 
- \langle \tilde r_i \rangle_{>\theta}^2)(\langle \tilde r_j^2 \rangle_{>\theta} 
- \langle \tilde r_j \rangle_{>\theta}^2)}}, 
\ee 
 where the subscript $> \theta$ means that both normalized returns are larger than  
 $\theta$, and $\tilde r_i$ are normalized centered returns.  
 The negative exceedance 
 correlation $\rho_{ij}^-(\theta)$ is defined similarly, the conditioning being now  
 on returns smaller than $\theta$. 
  We have plotted the average over all pairs of stocks $\rho^+(\theta)$ 
  for positive $\theta$ and $\rho_-(\theta)$ for negative $\theta$, both  
  for empirical data and for surrogate data generated according to a  
  {\it non-Gaussian} one factor model Eq. (\ref{eq2}), where both the market 
  factor and the idiosyncratic factors have fat tails compatible with empirical 
  data \cite{QF}. Note that empirical exceedance correlations {\it grow} with $|\theta|$  
  and are strongly asymmetric. For a Gaussian model, $\rho^{\pm}(\theta)$ would have  
  a symmetric tent shape, i.e. it would decrease with $|\theta|$ ! 
   
  In conclusion, most 
  of the downside exceedance correlations seen in Fig.\ 4 can be 
  explained if one factors in properly the fat tails of the 
  unconditional distributions of stock returns and the skewness of the 
  index \cite{QF}, and does not require a specific correlation increase mechanism.

  \begin{figure} 
  \psfrag{xmain}[ct][ct]{\huge time (trading day)} 
  \psfrag{ymain}[cb][cb]{\huge $\cal V$(t)}   
  \psfrag{xinset}[ct][ct]{ time (trading day)} 
  \psfrag{yinset}[cb][cb]{ ACF} 
  \centerline{\epsfig{file=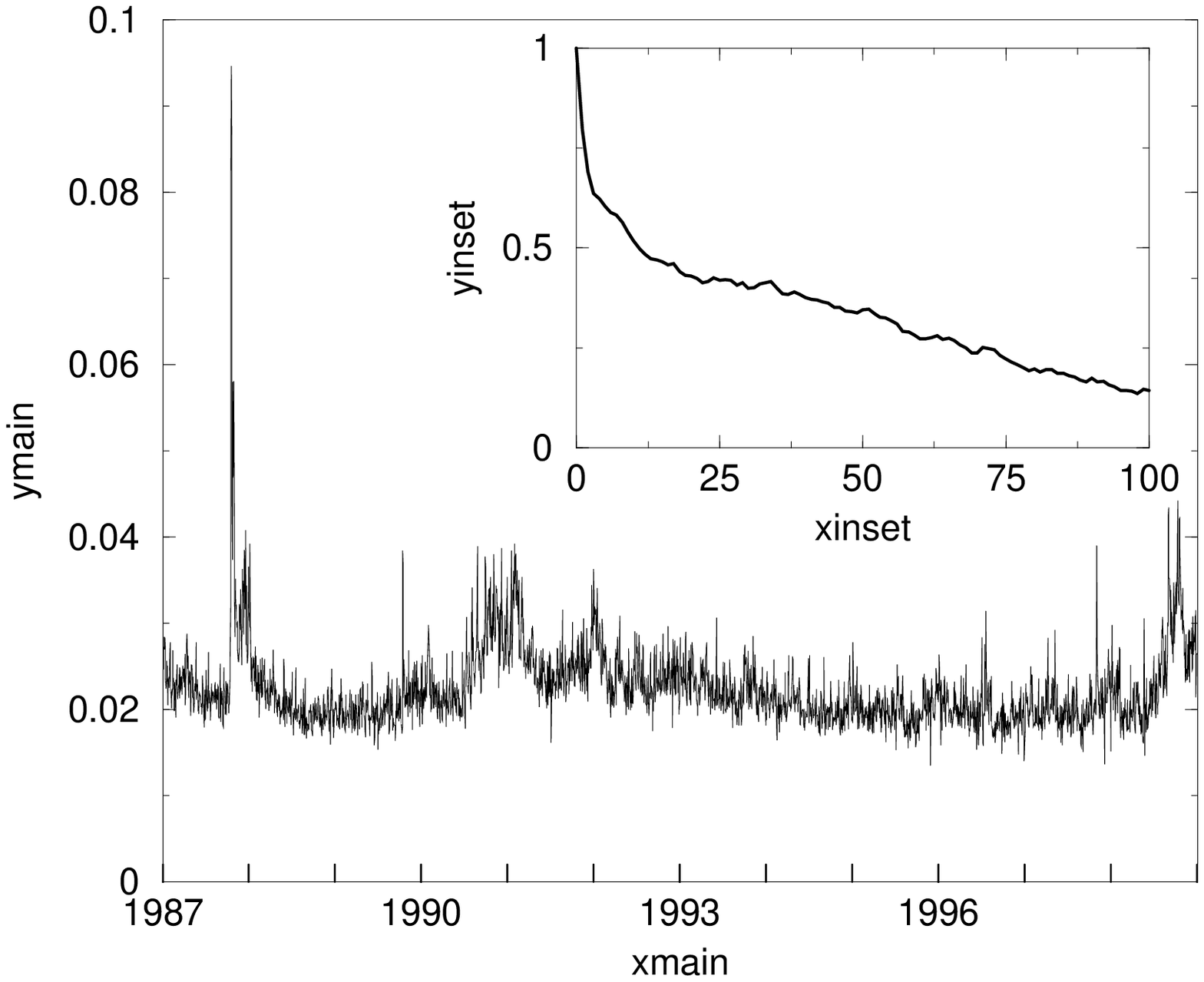,height=0.80\textwidth}} 
  \caption{Time evolution of the daily variety ${\cal V}(t)$ of 1071  
  NYSE stocks continuously traded from January 1987 to December 1998. 
  The time evolution presents slow dynamics and several bursts.  
  The highest peak is observed at and immediately after the Black Monday. 
  The highest value corresponds to the day after Black Monday. 
  In the inset we show the autocorrelation function (ACF) of ${\cal V}(t)$. 
  The autocorrelation has a slow decay in time and is still as high as  
  $0.15$ after $100$ trading days.} 
  \label{fig:variety} 
  \end{figure} 
 
  \begin{figure} 
  \psfrag{mean}[ct][ct]{\huge $r_m$} 
  \psfrag{variety}[cb][cb]{\huge $v$} 
  \centerline{\epsfig{file=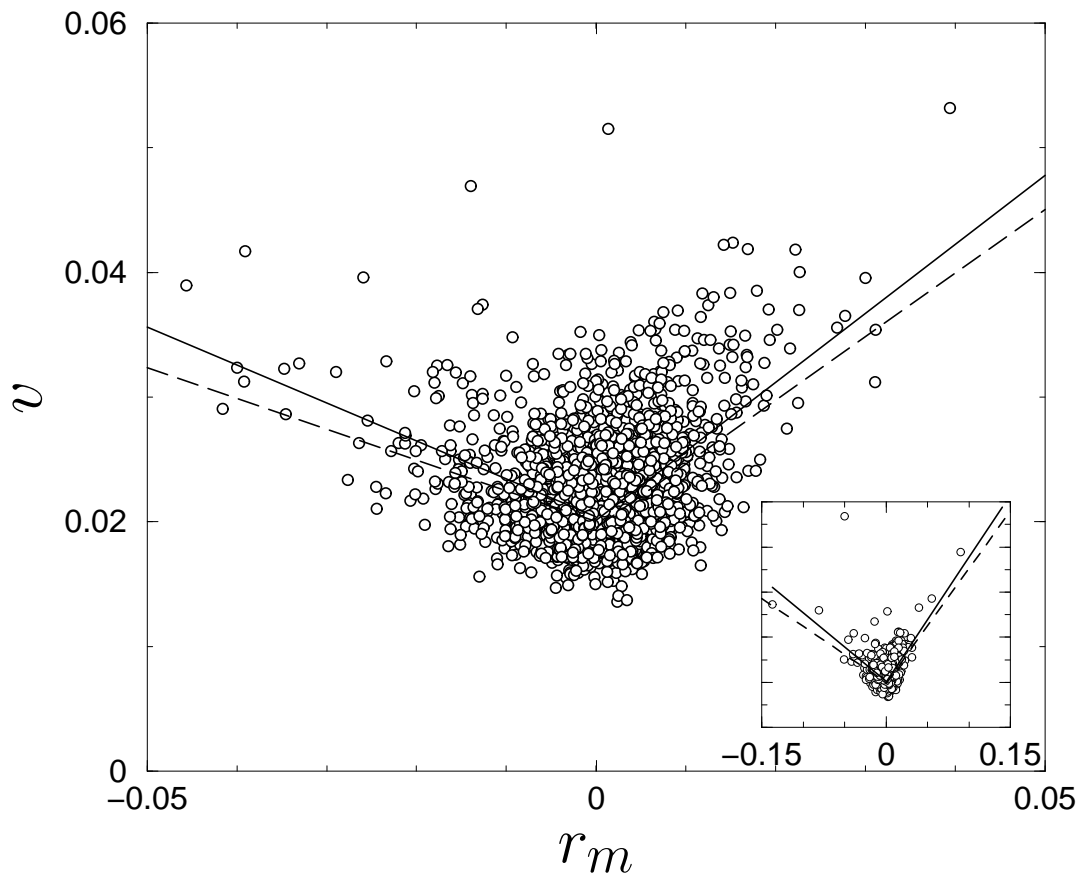,height=0.80\textwidth}} 
 
  \caption{Daily variety $v$ of idiosyncratic terms of the 
  one-factor model (Eq. (3)) as a function of  
  the market average $r_m$ of the 1071 NYSE stocks continuously 
  traded from January 1987 to December 1998. Each circle refers to  
  one trading day of the investigated period. In the main panel 
  we show the trading days with $r_m$ belonging to the interval  
  from $-0.05$ to $0.05$,  
  whereas in the inset we show the whole data set including five  
  most extreme days.  
  The two solid lines are linear fits over all days of positive 
  (right line) and negative (left line) market average. 
  The slope of the two lines are $+0.55\pm 0.02$ (right) and 
  $-0.30\pm 0.02$ (left). The tick distance in the  
  ordinate of the inset is equal to the one of the main panel. 
  The two dashed lines are linear fits obtained with a robust 
  local M-estimates minimizing the absolute deviation.  
  The slope of the two lines are $+0.51$ (right)  
  and $-0.25$ (left). These values are quite close to the 
  previously obtained ones, showing that the role of outliers is 
  minor.} 
  \label{fig:variety_idio} 
  \end{figure}

  \begin{figure} 
  \psfrag{xaxis}[ct][ct]{\huge $|r_m|$ (\%)} 
  \psfrag{yaxis}[cb][cb]{\huge $\cal C$} 
  \psfrag{legend1}[l][l]{\small Empirical data} 
  \psfrag{legend2}[l][l]{\small One factor model} 
  \centerline{\epsfig{file=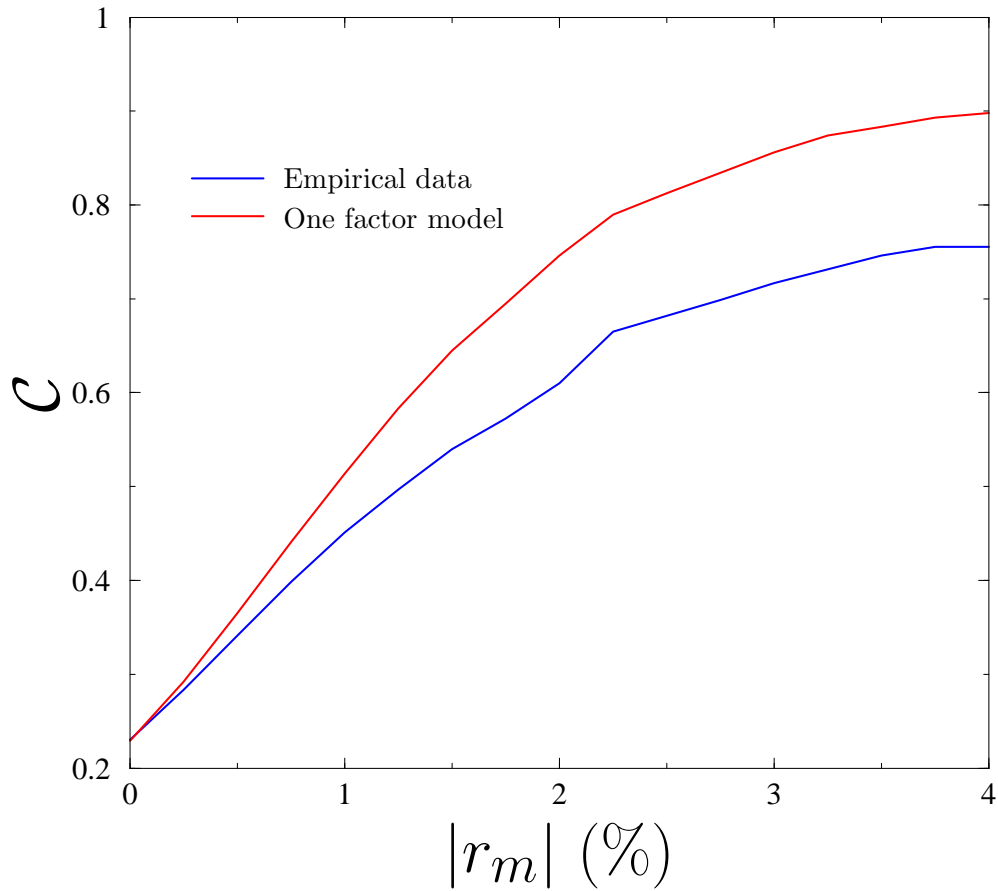,height=0.80\textwidth}} 
  \caption{Correlation measure $\cal C$ conditional to the 
  absolute market return to be larger than $|r_m|$, both  
  for the empirical data and for surrogate data using a (non 
  Gaussian) one-factor model \protect\cite{QF}. Note that both show a  
  similar apparent increase of correlations with $|r_m|$. This effect is 
  actually  
  overestimated by the one-factor model with fixed residual volatilities. 
  $|r_m|$ is in percents.} 
  \label{fig:condcorr} 
  \end{figure} 
 
  \begin{figure} 
  \psfrag{xaxis}[ct][ct]{\huge $\theta$} 
  \psfrag{yaxis}[cb][cb]{\huge $\rho^{\pm}(\theta)$} 
  \psfrag{legend1}[l][l]{\small Empirical data} 
  \psfrag{legend2}[l][l]{\small Surrogate data} 
  \centerline{\epsfig{file=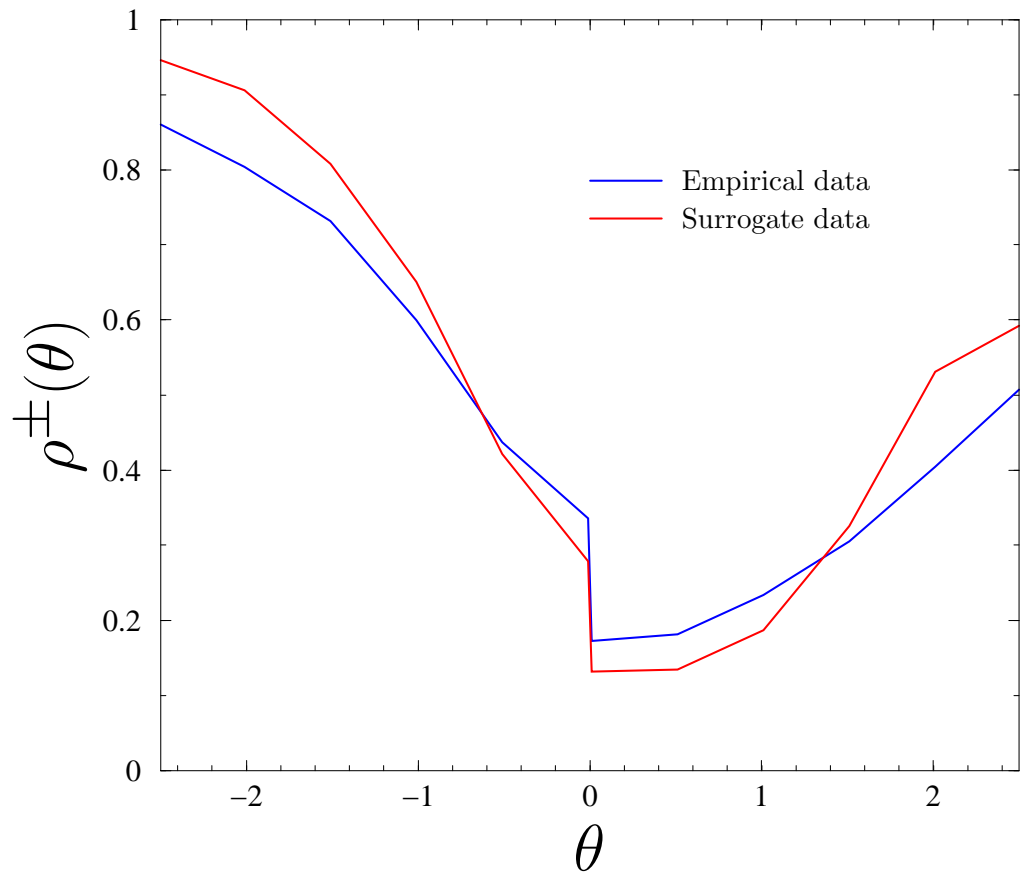,height=0.80\textwidth}} 
  \caption{Average exceedance correlation functions between all pair of stocks  
  as a function of the level parameter $\theta$, both for real data and  
  the surrogate non Gaussian one-factor model.} 
  \label{fig:exceedcorr} 
  \end{figure}

  \begin{figure} 
  \psfrag{mean}[ct][ct]{\huge $r_m$} 
  \psfrag{mean-median}[cb][cb]{\huge $\cal A$} 
  \centerline{\epsfig{file=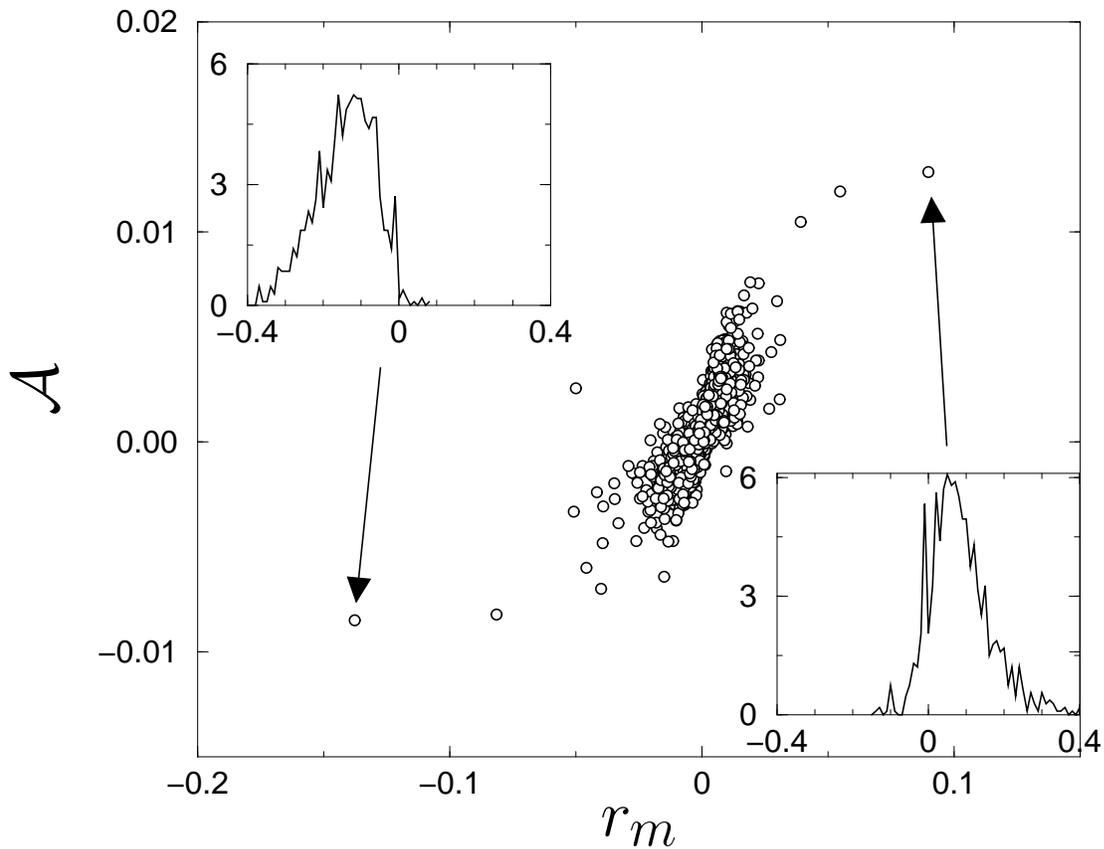,height=0.80\textwidth}} 
  \caption{Daily asymmetry ${\cal A}$  
  of the probability density function of daily returns of a set 
  of 1071 NYSE stocks continuously 
  traded from January 1987 to December 1998 as a function of  
  the market average $r_m$. Each circle refers to  
  one trading day of the investigated period. 
  In the insets we show the probability density function of daily returns 
  observed for the two most extreme market days of the period investigated. 
  Specifically, the left inset refers to Black Monday (October 19th, 1987) 
  and the right inset refers to October 21st, 1987. The negative  
  (left inset) and positive (right inset) skewness of the 
  distribution is clearly seen in both cases.} 
  \end{figure} 
 
  \end{document}